# IBM PC running Solaris x86 as an affordable NFS file server platform for a workgroup[*‡]

## Sailing instructions for system administrators

### Alexandre (Sasha) Telnov [☽]

*University of California at Berkeley, Department of Physics*
*Lawrence Berkeley National Laboratory, Physics Division*

### Abstract

The possibility of using an IBM PC–compatible computer as an NFS file server for a moderate-size workgroup has been explored. A pilot project currently under way at LBNL has proven that this platform does indeed provide a viable alternative to commercially available file servers at a fraction of the cost without significant sacrifices in performance and reliability.

From a $1,500 worth of hardware, we have built a Solaris x86 file server with 19.2 GB of disk space, expandable to 112 GB for another $6,200, which has proven to be a worthy contender in performance to its proprietary counterparts. The server has not exhibited any software of hardware problems during the four months it has been in service. Overall, the platform appears to be well suited to the needs of an experimental particle physics workgroup. We recommend it for use in the BaBar collaboration.

This paper focuses on aspects of building a properly configured dedicated file server that are not adequately covered in the *Solaris Installation* and *System Administration* guides, such as choosing right hardware and performance optimization. Parts of it may also be of interest to prospective buyers of IBM PC-compatible systems and to Solaris SPARC administrators.

> *"... any scientist who devotes himself to the study of Solariana has the indelible impression that he can discern fragments of an intelligent structure..."*
>
> Stanisław Lem, "Solaris"

## Contents

---

[*] This work was in part supported by the U.S. Department of Energy under Contract DE-AC03-76SF00098.

[‡] This article is a slightly modified version of BaBar Note #404. If you are affiliated with BaBar, you should get the original note from the BaBar Note Repository. BaBar is the name of the international experimental particle physics collaboration building a detector for studying physics of B-mesons at the PEP-II $e^+e^-$ collider at the Stanford Linear Accelerator Center. For more information, see the BaBar detector home page at
`http://www.slac.stanford.edu/BFROOT/doc/www/bfHome.html`

[☽] E-mail: `avtelnov@lbl.gov`

## 1. Introduction

The large size of BaBar code and data files<sup>☺</sup> imposes new tough requirements on file storage and network capacity. Given the today's speed of network connections between participating institutions and SLAC, most BaBar collaborators doing code development and physics analysis have only two options: using computers at SLAC (or LBNL) — or having local copies of BaBar software and data files at their home institutions. The question is, what is the most efficient way to do computing "at home" staying within the limits of the today's tightening R&D budgets? Could the kind of hardware sold in your local computer store at least partially replace those expensive computer systems traditionally used in particle physics laboratories?

For most BaBar applications, the two most important factors determining computing productivity are the CPU power and availability of disk storage for large data files. Since most BaBar applications are CPU-bound and do not require high I/O rates, the most logical solution is to distribute computing jobs across an array of (possibly booting from an OS server, possibly monitor-less, possibly multiprocessor) workstations (clients) with fast CPUs while keeping all code and data files on a small number of dedicated NFS file servers. Among advantages of this approach are a high degree of scalability in computing power, storage capacity *or* both, better performance under heavy loads, and the opportunity to significantly decrease the total cost by optimizing component selection and upgrading, adding more, or replacing clients and disks/servers when necessary instead of investing in expensive general-purpose servers (small mainframes, if you will), which are usually expandable, but not easily upgradable to faster CPUs and RAM.

For a variety of reasons, commercially available UNIX platforms, such as Alpha/Digital UNIX or UltraSPARC/Solaris, still have an edge over x86-based ("IBM PC-" or "Intel-compatible") computers running UNIX (Solaris x86, Linux, FreeBSD, etc.) or Windows NT. It is their superior floating-point performance, availability of commercial compilers and software packages used in the BaBar code, technical support from the manufacturer, etc. Still, it probably makes sense to preserve x86 as a possible option for the future, as the gap in performance (and price) continues to contract. Further research in this direction is necessary.

However, when it comes to disk storage, the situation is quite different. While DEC, Sun Microsystems, Silicon Graphics and other companies provide a wide selection of file server and disk storage solutions, they do not seem to be tailored to the needs of a moderate-size particle physics workgroup. Most of them are designed for mission-critical cutting-edge I/O applications, so they implement such features as hardware RAID arrays of hot-swappable and hot-pluggable wide UltraSCSI drives. These features are not really necessary for our purposes, since most files lost in the event of a serious file system crash are easily replaceable, only users' home directories and source code have to be located on backed-up file systems.

Most commercial file servers are picky about which drives they would accept, and which they would not. For example, the DEC StorageWorks 9.1 GB[1] narrow SCSI-2 (10 MB/s interface bandwidth) drive for DEC servers and disk arrays (3.5", 7200 RPM, 8 ms average seek time, 5-

---

<sup>☺</sup> We are talking about 25-50 MB executables, half-a-gigabyte of source code and libraries per code release, and gigabytes upon gigabytes upon gigabytes of data.
[1] For marketing purposes, hard drive manufacturers use units of 1 "GB" ≡ 1,000,000,000 bytes, not 1 GB ≡ $2^{30}$ = 1,073,741,824 bytes.

year warranty) has a list price of $2499; the wide UltraSCSI (40 MB/s) version has a list price of for $2599 (in November 1997, these drives could be bought for approximately $2000)[2]. General-use, non-proprietary drives are much cheaper. For example, at the time this note is being written, Bay Area computer hardware retailers are selling the Quantum Fireball ST 6.4 GB narrow UltraSCSI (20 MB/s) drive (3.5", 5400 RPM, 9.5 ms, 128k cache, 3-year warranty) for about $350; Quantum, Seagate, IBM, Metropolis 9.1 GB wide UltraSCSI 3.5", 7200 RPM, 9 ms, 1MB cache, 5-year warranty drives go for less than $750; the Seagate Elite 23.2 GB wide UltraSCSI drive can be purchased for $1550. However, these drives cannot be used in DEC's file servers. EIDE disks traditionally used in IBM PC-compatibles are even cheaper: a brand-name 8.4 GB EIDE Ultra ATA (33 MB/s) disk with a 3-year warranty costs about $300.

Since most simulation and data analysis applications are not I/O-intensive, the server throughput does not have to be very high — capacity is more important. Disk and network I/O do not use floating-point instructions, so the server processor's FPU performance is irrelevant. In addition, SCSI controllers use direct memory access and do not put a heavy load on the CPU, so that it does not have to be very powerful. Since in most typical situations in UNIX NFS environments most caching is taken care of by the clients, the server need not have a lot of RAM. If the server is not to be used for anything else except serving file systems over NFS, the choice of the hardware platform and the operating system is relatively arbitrary, since it does not affect users. It is dictated by reliability and security considerations and by hardware, software and support costs.

The goal of the pilot project started at LBNL in October 1997 by the author of this paper is to answer our need for more scratch disk space by building a proof-of-principle dirt-cheap, yet fast, reliable, expandable and upgradable x86-based NFS server. The server has been up and running since October 15, 1997.

## 2. Why Solaris x86

When choosing the "flavor" of UNIX that is right for you, the following things have to be taken into consideration, in no particular order: reliability, performance, price, documentation (printed and online manuals, books), technical support (from the OS manufacturer, from discussion groups on Usenet and WWW, from you local Computer Support Group), availability of desired features (for a server, it would mean disk concatenation, stripping and mirroring, software RAID levels 0, 1, 5, etc.), software availability, security features. Which factors are more important depends on the system's purpose and the environment in which it will be used.

The x86 UNIX market is clearly dominated by variations of Linux[3] and BSD[4] (FreeBSD and NetBSD), which are free. Each of them is developed and maintained by a "large team of individuals", who get help from thousands of "computer enthusiasts", a.k.a. "hackers", from all over the world. The source code for these versions of UNIX is freely available, so that anyone can write a driver or a system utility — or try to find a loophole in the system's security. Both systems have excellent performance[5], online documentation and discussion groups, and most features of the latest commercial versions of UNIX. They are easy to install, especially over the network, since device drivers are usually more up-to-date than in the case of commercial UNIXes, and GNU utilities and development tools are included in the distributions. However, the lack of consistency and manufacturer support and potential security risks make free versions of UNIX poorly suited for the environment of a national laboratory or a big corporation.

---

[2] See `http://www.storage.digital.com/swrks/guide/pg_ddrv.htm`, `http://ww1.systems.digital.com/Ddic.nsf/HWStorageProducts`.

[3] See, for example, `http://www.redhat.com`, `http://www.debian.org`, `http://www.linux.org`, `http://www.linuxhq.com`, `http://www.li.org`.

[4] See `http://www.freebsd.org`, `http://www.netbsd.org`.

There are three major commercial UNIX implementations that run on IBM PC–compatibles: Sun's Solaris x86[6], SCO UnixWare[7], and BSDI's BSD/OS[8]. Out of these three, Solaris x86 is the most popular one. Virtually all features available in Solaris SPARC are also available in Solaris x86. The high degree of source code compatibility between the SPARC and the x86 versions, limited only by availability of commercial development tools, makes Solaris x86 an attractive choice for environments where SunOS has been used for a long time. Another factor is the Sun's reputation. The LBNL's Unix Systems Group is currently working on obtaining a site license for Solaris 2.6; it is going to be the first and the only supported x86 UNIX implementation at LBNL.

## 3. Hardware selection guide

We are using mainstream brand-name x86-compatible hardware, which is close to the leading edge in performance, but has been on the market for at least 6 month and therefore costs much less. In most cases, parts with desired features are available from multiple manufacturers, and in this case the choice depends on manufacturer's reputation, part's price and availability. If you decide to get an x86-based file server, it is probably best to order a custom-built system from a local computer store or a reputable company that specializes in servers and workstations built to meet specific customer requirements and sells direct to end-users. Local computer stores are often more flexible and offer equal or superior products and comparable service at a significantly lower price, while big companies usually offer more comprehensive warranty and service plans and have access to specialized components — but you may be limited to their product offerings and upgrade paths.

It is absolutely necessary to consult the most recent *Solaris x86 Hardware Compatibility Guide* before choosing add-on cards, such as video and Ethernet cards, SCSI and additional EIDE controllers. Avoid on-board video and SCSI controllers, unless the *Guide* specifically lists the mainboard in question — or have them disabled. Other than that, any standard mainboard should work. The choice of the CPU, EIDE or SCSI drives, the CD-ROM drive and other components should not cause any compatibility problems. The latest versions of the *Guide* and *Driver Update* floppies can be found on the Solaris web site.

---

[5] See K. Lai, M. Baker, *A Performance Comparison of UNIX Operating Systems on the Pentium.* Proceedings of the 1996 USENIX Conference, San Diego, CA, January 1996. The authors compare performance of Linux, FreeBSD, and Solaris x86 in a wide range of benchmarks. The article can be downloaded in PostScript format from `http://gunpowder.stanford.edu/~mgbaker/publications/index.html`.

[6] See `http://www.sun.com/solaris`

[7] Novell transferred UnixWare to SCO in December 1995. See `http://www.sco.com/unixware`

[8] See `http://www.bsdi.com`.

For reference, here is the hardware configuration of the pilot machine and some comments that might help understand what is up on the market of x86-compatible hardware. Prices shown are what we paid in September 1997, followed [in square brackets] by typical prices of the same components in February 1998. Your choice of parts will probably be very different from ours.

- **Case and power supply:** Full Tower ATX-formfactor with a 250W power supply, six 5.25" and four 3.5" bays, $121 [$80]. The ATX mainboard formfactor is gradually replacing the old AT standard, so we picked ATX — just in case if some time in the future, when ATX completely replaces AT, we decide to upgrade the mainboard. Keep in mind that the sum of peak power consumption rates of all system's components has to be less than about 110% of the power supply. Typically, it's about 20W per drive plus 40W for everything else, but you better consult the manuals. Cases with more bays and more powerful (up to 600W) redundant power supplies are available.

- **CPU:** Cyrix/IBM M1 6x86-PR200+ (150 MHz, 75 MHz system bus x 2) with a fan, $81 [$55]. This CPU has better integer performance than Pentium-200 and is pretty close in performance to the top-of-the-line Pentium II CPU's – for a fraction of the price. Just like Intel Pentium, Pentium MMX, AMD K6 and IDT WinChip CPUs, this is a Socket 7 processor (Pentium II uses Slot 1, which is not licensed by Intel to other chip manufacturers; Pentium Pro's Socket 8 design is dead). Cyrix/National Semiconductor, AMD and IDT are working on further improving internal architecture and clock speed of their Socket 7 CPUs (Cyrix/IBM 6x86MX-PR233 [$140] and AMD K6-PR233 [$160] already beat Pentium II-233 [$270] in performance[9], 300+ MHz models are expected in Q2 '98), so the CPU that we have now can be upgraded with a more powerful one, if necessary.

    The only real advantage of the Intel Pentium II processor is support of the Intel's proprietary multi-processing standard, APIC, by chipset and motherboard manufacturers. Having dual processors can noticeably improve server's performance, especially if there is a lot of EIDE disk activity (see below), but it will also require obtaining a Solaris Server license, which is much more expensive than a Solaris Desktop license. If you do get a Server license and plan on having a several thousand dollars worth of disks, it is probably worth paying an extra $700 for a Dual Pentium II-233 machine.

- **Mainboard:** QDI Titanium IIIB ATX-formfactor jumperless, i430TX chipset, 512k L2 cache, $107 [$100]. The Intel 430TX chipset is the only Intel's Socket 7 chipset supporting Ultra ATA 33MB/s, a.k.a. Ultra DMA, the newest extension of the EIDE hard disk and CD-ROM standard. It only supports 128 MB of RAM and 512k of L2 cache (only the first 64 MB is L2-cached). New SiS 5581/2, SiS 5597/8 and VIA chipset-based mainboards, such as Asus SP97-V or Abit PD5N, support up to 512MB of RAM, all of it cached, and 2MB of L2; unfortunately, they were not available when our system was ordered. Future Socket 7 CPUs will work at 2.5V, 2.4V or 2.1V, unlike today's 2.8V or 2.9V, so if you might consider upgrading the CPU without changing the mainboard, make sure that the motherboard supports those voltages. Future CPUs will work at a higher than standard system bus frequency (75 or 83 MHz instead of 66 MHz), so getting a SiS-based mainboard with asynchronous 33 MHz PCI Bus and official support for those higher frequencies could be a wise thing to do. Also decide what kind of memory to use (EDO DRAM or SDRAM, see below) and whether you want memory with parity or ECC (error control and correction).

- **Memory:** Four 32MB (8x32Mb) 60ns EDO DRAM non-parity tin-plated 72-pin SIMMs, $120 [$55] each. There are two formats of memory modules on the market today: the new 168-pin DIMM format, and the old 72-pin SIMM format (since Pentium-class processors have a 64-bit memory bus, you must always have a pair of identical SIMMs). DIMMs will probably push SIMMs out of the market in three or four years. Motherboards sold today may support 2, 4, or 6 SIMMs, 2 or 3 DIMMs (3.3V, unbuffered), or a combination of them. Currently, the maximum capacity of one memory module is 128MB (32x32Mb) for SIMMs and 128MB (16x64Mb) for DIMMs. Both SIMMs and DIMMs are available in the ECC and parity (x36Mb for SIMMs, x72Mb for DIMMs) versions, but only DIMMs are available in the SDRAM (10ns) version. The more expensive SDRAM memory has a twice-higher bandwidth than EDO DRAM and will work at 100MHz. However, in most cases benefits from using SDRAM and/or ECC or parity memory are not significant. Since 128MB of RAM is enough for a dedicated file server, our choice of memory configuration seems to be well justified.

---

[9] See, for example, `http://www.pcworld.com/news/daily/data/0597/test_report-1b.html`

- **Hard drives:** Three Quantum Fireball ST 6.4GB 3.5", EIDE Ultra ATA 33MB/s, 14MB/s sustained hardware disk-to-cache transfer rate, 5400 rpm, 9.5 ms AST (15.1 ms latency), 128k cache, $330 [$250] each. Pentium mainboards have two on-board EIDE controllers. Each controller can support one or two EIDE devices, such as hard, CD-ROM or Zip drives. EIDE hard drives have less on-drive electronics, and that is one of the reasons why they cost less than SCSI drives. The narrow UltraSCSI (20 MB/s maximum cache to controller transfer rate) version of Quantum Fireball ST 6.4 GB with the same 128k of cache costs $120 more. A typical SCSI drive, however, has 512 kB or 1 MB of cache, and higher spindle rotational speed (7200rpm or 10000rpm). It does not make much sense to equip a file server with very fast drives, since for large files the disk I/O rate attainable with 5400 rpm drives is already higher than the NFS over TCP/IP over Ethernet throughput, which is about 1 MB/s for 10Base-T Ethernet and 3-4 MB/s for 100Base-TX, and for small files the files-per-second rate depends primarily on the disk latency, which is not less than 10.5 ms even for the most expensive hard disk drives.

    The choice between EIDE and low-end SCSI drives is more difficult. The interface speed does not really matter: the real I/O rate you can get will never get close even to the sustained transfer rate stated in the drive's manual. SCSI drives are a little more expensive, but they are easier to add to a system. Unlike SCSI, EIDE controllers do not support Bus Mastering DMA under Solaris x86. Every time an EIDE controller wants to access the drive's I/O buffer, it interrupts the CPU. As a result, EIDE I/O in Solaris x86 is very CPU-intensive. We have found that for our configuration the maximum local disk I/O rate is CPU-bound at about 4.7 MB/s for TMPFS and 3.9 MB/s for UFS. Overall, it probably makes sense to have at least one EIDE hard drive for the OS to boot from; it makes no sense to have more than 3 EIDE hard drives (Solaris x86 supports up to 8 EIDE devices). If we decide to expand our server, we will buy an UltraSCSI controller and four large low-end SCSI drives.

- **CD-ROM drive:** TEAC 24x EIDE, $95 [$60]. Any standard EIDE CD-ROM drive is OK. A CD-ROM drive is optional, since the Solaris installation program loaded from floppies can NFS-mount the Solaris CD from a remote host.

- **Ethernet Card:** 3Com Fast Etherlink XL 10/100Base-T(X) (3C905 TX), $89 [$70]. This Ethernet card is not supported by the original release of Solaris 2.5.1, so you have to download the latest driver update boot and distribution floppies before installing Solaris.

- **Video Card:** Diamond Stealth 3D 2000 2MB PCI, $43. The cheapest video card we could find. Does 1024x768 @ 8 bit, enough for a server's console. Required video driver update floppies.

- **Mouse:** 3-button PS/2-style (6-pin mini-DIN) mouse, $12. All ATX and some AT mainboards support mice with mini-DIN connectors. A serial port mouse would work too.

- **Keyboard:** 104-key PS/2-style (6-pin mini-DIN) keyboard, $24. ATX mainboards have only a mini-DIN keyboard connector, AT mainboards have only a 5-pin DIN keyboard connector.

- **Monitor, floppy drive:** no comments.

## 4. Operating system and software installation

Most necessary information on Solaris installation and configuration can be found in the manuals that come with the OS or can be found on the WWW[10]. Here I briefly describe our system's configuration and some details you should be aware of before beginning installation.

We use Solaris x86 version 2.5.1, the latest one available at LBNL in October 1997 (the newest version of Solaris x86 is 2.6), with *Driver Update 10* and *Video Driver Update 10*[11]. In our case, it was necessary to use the three *Driver Update 10 Boot* and the three *Driver Update 10 Distribution* floppies instead of the original *Boot* floppy, since drivers for EIDE Ultra DMA drives and the Etherlink XL card were not available when Solaris x86 version 2.5.1 was released. Since our video card was not supported by the original Solaris release either, the OS installation process was text-based; *Video Driver Update 10* had to be installed later to enable X–Windows. Other than that, Solaris installation is pretty straightforward.

The OS is installed on the master drive on the primary EIDE controller (`c0d0`); the slave drive on the secondary controller (`c1d1`) is the CD-ROM; the other two hard drives are `c0d1` and `c1d0`[12]. On the boot drive (`c0d0`), 305 MB is given to the root (`/`) partition[13] (`c0d0s0`), 258 MB – to `swap` (`c0d0s1`), and 1906 MB – to `/usr` (`c0d0s6`). The rest of disk space (about 15 GB) is unified in a single Solstice DiskSuite 4.0[14] metadevice[15] (`/dev/md/dsk/d0`), which is served

---

[10] Solaris manuals in PDF and HTML formats can be found at
`http://www.ccd.bnl.gov/~stange/sundoc.html`, `http://ultra.cto.us.edu.pl:8888/`,
`http://www.kom.auc.dk/EDB/doc/answerbooks/`, `http://faui40f.informatik.uni-erlangen.de:8888/ab2` and many other locations.

[11] You can get images of the latest boot and driver update floppies for all versions of Solaris x86 from
`http://access1.sun.com/drivers/driverMain.html`. The images can be transferred to 3⅓" 1.44 MB floppies either using `dd` on UNIX, or by using a freeware utility called `dcopy`
(`http://online04.lbl.gov/~telnov/dcopy.exe`) on DOS or Windows.

[12] Note that we connected the two drives that will be loaded most heavily to two different EIDE controllers. Since the controller interface speed (33 MB/s) is much larger than the real life data throughput one could expect from two EIDE drives, this gives only a slight increase in performance compared to connecting both of them to the same controller.

[13] In Solaris manuals, the terms "partition" and "slice" refer to the same thing and are used interchangeably.

[14] The Solaris Solstice Tools, such as DiskSuite and AdminSuite, are not included in the Solaris Desktop distribution, so you must get Solaris Server if you want to use advanced disk management (mirroring, concatenation, stripping, RAID, etc.) The Solstice DiskSuite CD contains both SPARC and x86 binaries.
Besides the bundled software, the only difference between Solaris Desktop and Solaris Server is the license it comes with. The Desktop license allows running Solaris on a single-CPU desktop with no more than 2 users logged in at a time, while the Server license permits server activities on a machine with up to 4 CPUs and up to 5 users.

[15] (the following are some very technical details of our DiskSuite setup, which are here to give you an idea of how a complex metadevice may be configured). Solaris x86 version 2.5.1 interprets master drives on both EIDE controllers (`c0d0` and `c1d0`) as having less than 1024 cylinders by dividing the number of physical cylinders and multiplying the number of read/write heads by some number (usually by 17). So, from the Solaris' point of view, `c0d0` and `c1d0` have 255 heads, 780 tracks and 16065 blocks per cylinder, while `c0d1` has 15 heads, 13325 tracks and 945 blocks per cylinder. After taking into account that one DiskSuite metadevice state database replica is 1034 blocks in size by default and always occupies an integer number of cylinders, we assigned all available space on `c1d0` to one partition (`c1d0s5`, cylinders 3-777) and divided `c0d1` into two partitions (`c0d1s3`, cylinders 3-161, and `c0d1s5`, cylinders 162-13322). Now if we put 2 replicas onto `c0d1s5` and 3 replicas onto `c1d0s5`, the number of data blocks available to DiskSuite metadevices on `c0d1s5` and `c1d0s5` is the same (!), so we can use striping on them without any loss of

over NFS. **IMPORTANT:** *Configuring DiskSuite the first time is not straightforward, so be sure to read and understand the Solstice DiskSuite Administration Guide, the Solstice DiskSuite User Guide or/and the Solstice DiskSuite Reference before trying to create any metadevices — or, better yet, before partitioning the hard drives.*

Unfortunately, the Solaris distribution comes without compilers, development tools and many useful utilities, so we had to download them from the Net. There are many archives where you can get Solaris x86 binaries of the most popular GNU software. In some cases, we had to build the binaries ourselves. Here is the list of software we have installed on our server: `gcc`, `g++`, `libg++`, `gdb`, `gzip`, `emacs`, `less`, `tcsh`, `zsh`, `bash`, `zip`, `unzip`, `top`, `traceroute`.

We chose not to use NIS for user management. In principle, it is enough to create local groups complying with the Lab-wide assignment of group names and GIDs used by members of our workgroup and change the exported directory's permissions (`cd /export/x86serv ; chown root:bbrgrp . ; chmod 775 .`). In order to be able to see usernames instead of UIDs, users must be registered in the local `/etc/passwd` file. An option in the `/etc/shadow` file, which contains encoded passwords and other security-related information, can be used to disable logins by ordinary users (remember that `/etc/shadow` and `/.rhosts` must have their access permissions set to `-r--------`).

In order for `gmake` and other programs to operate properly, clocks on all systems within a network must be synchronized. Add something like "`59 23 * * * /usr/bin/rdate unixhub.slac.stanford.edu`" to `/var/spool/cron/crontabs/root`.

## 5. Performance tuning[16]

> *Data: Is there a problem with the engines?*
> *Geordi: No.*
> *Data: Then why are you stabilizing the EPS conduit?*
> *Geordi: Just trying to get a slightly higher power conversion level.*
> *Data: But that would not affect the engines in any way!*
> *Geordi: I know that, Data. It's not the point… I am just trying to get a higher conversion level, that's all.*
> *Data: Why?…Are you in competition?…*
> *Geordi: Actually, it's more a matter of personal pride. These are _my_ engines.*
>
> *Star Trek: The Next Generation, "Power of Nature"*

The general opinion on tweaking for increased performance is that it is debatable whether any significant improvement would be worth the effort. There is no magic bullet, a secret tweak to a kernel variable that will miraculously make the whole machine run much faster. It is rare for this kind of change to make a difference that can be reliably measured at all. There are, however, a few common kernel tuning variables that can be adjusted to reflect the *specific purpose* of our machine, significantly improving its performance as a file server. There is little systematized information available on tuning kernel parameters; adjustments beneficial in one version of the

---

disk space. The 73 MB of space on `c0d1s3` can be used for DiskSuite UFS logging, be concatenated to the stripe or be used for something else. In our case, `/dev/md/dsk/d0` is a *concatenated stripe* made of `c0d1s5` (2 replicas) and `c1d0s5` (3 replicas) striped with the interlace of 48 kB, `c0d0s5` (2 replicas) and `c0d0s7` (no replicas).

[16] Most recommendations in this chapter apply both to x86 and SPARC Solaris.

operating system may be unnecessary or harmful in later releases; some advises found on the WWW or Usenet are nothing more than folklore. Deciding when a tweak is useful and what the value should be is complex and requires testing. Since one of the purposes of this paper is to help guide a future system administrator through the whole process of configuring a Solaris x86 fileserver, we conducted research and worked out a set of recommendations that make sense — but there is no guarantee that our recommendations are flawless. They are a mixture of theory[17], informed guesswork and extensive testing.

Solaris provides a convenient to use system activity reporter utility `sar`, which gathers statistics about various kinds of system activity and can be run either interactively — or in the Automatic Data Collection mode, in which case it keeps a database of system activity statistics for the last 30 days. For example, `sar -a` displays file access statistics, `sar -b` displays buffer activity statistics, `sar -d` displays disk activity, and `sar -u` displays CPU utilization. Other useful system statistics utilities are `vmstat`, `iostat`, `netstat`, and `mpstat`. These utilities provide valuable information that helps pinpoint problems and bottlenecks and improve performance, if possible, by tuning kernel or device driver parameters. When tuning performance, it is a good idea to have a working copy of the kernel subtree and the `/etc/system` file, which can be used to boot the system in the event of a failed test[18].

A few words about the performance that can be expected from a file server. For large files, it is limited either by the networking speed or by disk I/O throughput (unless the server is repeatedly serving same files that fit into its file cache). For 10 Mbit/s Ethernet, server's throughput usually gets close to the wire speed (bandwidth) of 1.25 MB/s; the difference comes primarily from TCP/IP and Ethernet headers and checksums and from TCP reception acknowledgements, which constitute at least 12% of network traffic. For 100 Mbit/s Ethernet, the server's CPU speed and network interface performance become important. Discussion of Fast Ethernet performance deserves a separate article, so we will limit ourselves to noting that usually a server on 100 Mbit/s Ethernet can put through a lot less than its nominal bandwidth of 12.5 MB/s, primarily due to the NFS overhead, the fact that in Fast Ethernet the maximum amount of data per transmission (MTU) is still limited to 1500 bytes, and its latency (the round-trip time reported by `ping -s`) is just a little better than that of 10 Mb/s Ethernet (sub-millisecond vs. 1 ms).

For small files, the most important performance-limiting factor is the disk latency, which consists of two parts: the Average Seek Time (AST) and the rotational delay. The AST is the average time it takes to move the read/write heads to the specified track. It is 9.5 ms for our drives and can be as small as 7.5 ms for the most advanced HDDs manufactured today. The rotational delay is approximately one half of the inverse of the disk's rotational speed. For our disks, it's 5.6 ms (5400 rpm); for the best HDDs you can get it's 3.0 ms (10000 rpm). The AST and the rotational delay add linearly, so our drive's latency is approximately 15.1 ms; top-of-the line HDDs have a latency of 10.5 ms. As you can see, the difference is not dramatic. When we are reading small files that are located in random places all over the disk, the time per file is equal to the disk's latency plus the CPU overhead, provided that we already know where each file is located, i.e. if all relevant inodes are in the inode cache and all relevant directory entries are in the DNLC. If they are not, they have to be read from the disk as well, adding a few milliseconds to the file access time. Since processes usually do not request the next file until the previous one has been read, the drive/controller cache does not help. The only reason why we see small file I/O rates better than 1/15.1 ms = 66 files/s in our tests is because the files are located close to each other and the seek time is small. NFS adds more overhead, since it takes at least two round-trips to open a remote file. Writing is always slower than reading, especially for small files, because the CPU has to take care of file allocation.

---

[17] See "Sun Performance and Tuning: SPARC and Solaris" by Adrian Cockcroft, SunSoft Press PTR Prentice Hall, 1995. There are also a few Sun's White Papers that can be found on the WWW ("Sun Performance Tuning Overview", 1993, etc.) These sources are a little outdated, but in conjunction with discussions in SunWorld Magazine (`http://www.sun.com/sunworldonline`) they are a valuable source of information.

[18] `cp -r /platform/i86pc/kernel /platform/i86pc/goodkernel; cp /etc/system /etc/system.good`

The following network tuning has been performed: we changed the defaults for the maximum values of the TCP transmit and receive windows. The maximum size of both parameters is 65536 bytes. The necessary size of the transmit and receive windows depends on the bandwidth-delay product, which is the connection speed multiplied by the time it takes to transmit a data fragment and get a reception acknowledgement. This network latency time can be measured with the `ping` command. For Fast Ethernet LANs, the latency usually is less than 1 ms; for 10 Mb/s Ethernet LANs it is about 1 ms and usually stays below 10 ms for LANs connected through routers. The bandwidth-delay product is therefore almost always less that 10 kB, except for high-speed high-latency connections such as international fiber-optical or satellite links. It therefore does not make sense to set the maximum TCP `xmit` or `recv` window sizes to the maximum. Another parameter that we have adjusted is the TCP listen backlog `tcp_conn_req_max`. This parameter must be set to the maximum of 1024 on machines that have to handle *very* many open TCP connections, such as Web servers. In our case 64 should be enough. One more thing: be sure to have IP forwarding turned off: `ndd -set /dev/ip ip_forwarding 0`.

| Parameter name | Default size | New size | Line to add to `/etc/rc2.d/S69inet` |
|---|---|---|---|
| `tcp_xmit_hiwat` | 9216 | 49152 | `ndd -set /dev/tcp tcp_xmit_hiwat 49152` |
| `tcp_recv_hiwat` | 41600 | 49152 | `ndd -set /dev/tcp tcp_recv_hiwat 49152` |
| `tcp_conn_req_max` | 32 | 64 | `ndd -set /dev/tcp tcp_conn_req_max 64` |

Solaris is shipped with the number of concurrent requests that the NFS server daemon can handle set to 16. This is okay for a desktop on 10 Mb/s Ethernet, but too little for a dedicated NFS server on 100 Mb/s Ethernet or FDDI. You definitely have to increase this number, but by how much depends on how many NFS clients are served. Even setting it to a few hundred will not hurt. Change the following line in `/etc/init.d/nfs.server`: `/usr/lib/nfs/nfsd -a 64`.

It has already been mentioned that Solaris x86 2.5.1 does not support the Bus Mastering mode for EIDE drives, so the CPU has to take care of all disk I/O operations. This makes EIDE I/O very CPU intensive. EIDE controllers support some amount of buffering, with the purpose of interrupting the host only when an entire buffer full of data has been read or written, instead of interrupting for each sector. Some controllers hang when buffering is used, so the `ata` driver has a parameter used to reduce buffering[19]. The `drive#_block_factor` parameters in `/platform/i86pc/kernel/drv/ata.conf` can be increased to improve performance. The allowed values are 0x1 (default), 0x2, 0x4, 0x8 and 0x10. We chose not change the default values for the boot drive and the CD-ROM. **Caution!** Changing these parameters might cause your system to lock up during booting, so create a duplicate of the kernel before experimenting.

---

[19] See the `man` pages for the `ata` driver for more information.

| Controller | Drive | Parameter name | Default | New value |
|---|---|---|---|---|
| Primary IDE controller | HDD #1 (boot) | `drive0_block_factor` | 0x1 | 0x1 |
| | HDD #2 | `drive1_block_factor` | 0x1 | 0x10 |
| Secondary IDE controller | HDD #3 | `drive0_block_factor` | 0x1 | 0x10 |
| | CD-ROM | `drive1_block_factor` | 0x1 | 0x1 |

When a new file system is created, Solaris tries to optimize sequential disk I/O by taking into account the rotational speed of the disk. The default is 3600 rpm. It can be changed: `newfs -v … -r 5400 … `. There is no need to change the default values of `rotdelay` (the `-d` option, defaults to 0) and `maxcontig` (the `-c` option, defaults to 7). These three options are relics of the time when CPUs and disks were much slower than today.

The `maxpgio` kernel parameter (see below), which sets the maximum number of disk page out operations per second, is set to 2/3 of the rotation rate in rps of the disk used for swapping, leaving some bandwidth for regular processes. The default is (3600/60)*(2/3) = 40; it can be increased for faster drives[20].

The rest of performance tuning involves adjusting kernel parameters related to caching the file system and the file access information. We have discovered that using defaults for virtual memory paging algorithms impairs filesystem cache performance for files larger than approximately 10 MB by prematurely recycling memory pages in the file cache. The complexities of the entire virtual memory system and the page replacement algorithm are beyond the scope of this article, and we will not go into details. This is not a big issue anyway since UNIX NFS clients do a great job caching remote disks, so a file read from the server once is likely not to be read by the same client again for a long time. We have been able to improve file cache performance by increasing `lotsfree`, the minimum amount of memory (in pages) on the free list that triggers the page-out scanner, from the default 1/64[th] of the amount of physical memory to 1/16[th], but this is probably not the right way to go since the problem seems to actually be a bug in the Solaris x86 2.5.1 kernel (the problem has not been observed on Solaris SPARC 2.5.1 and Digital UNIX 4.0 machines).

The most significant performance gain we have achieved is related to caching file access information. All information about files, except their names, is kept in parts of cylinder information blocks called *inodes*. The inode information about currently and previously opened files is kept in the inode cache, which is a part of the UFS metadata buffer cache. `ufs_ninode` limits the number of inactive inodes in the cache. If a file is opened again, its inode information will not have to be read from the disk, saving us a few milliseconds. The inode cache entries also provides the location of every file data block on disk and the location of every page of file data that are in the RAM. If an inactive inode is discarded, all of its file data in memory are also discarded, and the memory is freed for reuse (such events are reported by `sar -g` as `%ufs_ipf`). If the clients are requesting a lot of tiny files, such as source code, `ufs_ninode` has to be set high enough to keep all their inodes. Its default value for our configuration, 2232, is too small for a server. The optimal value of this parameter depends on how many files the server is expected to serve on a regular basis and the amount of RAM (see `bufhwm` below). We set `ufs_ninode` to 12288.

---

[20] Actually, the virtual memory algorithms in Solaris 2.4+ have been improved in such a way that disk page out operations are virtually nonexistent on dedicated NFS servers.

The Directory Name Lookup Cache (DNLC) stores directory information. A directory is a special kind of file that contains names and inode number pairs. The DNLC holds the name and a pointer to an inode cache entry. When a file is opened, the DNLC is used to figure out the right inode from the filename given. If the name is in the cache, there is no necessity to read the directory information from the disk. NFS clients hold a file handle that includes the inode number for each opened file, enabling each NFS operation to avoid the DNLC and go directly to the inode. Thus a large size of the DNLC cache (`ncsize`) is only beneficial if there are multiple clients accessing the same source code files. Another possible reason to have a large DNLC cache is improperly tuned NFS clients that run out of space for remote node information. In Solaris, the maximum number of rnodes kept by an NFS client is set by the tunable kernel parameter `nrnode`, and defaults to approximately $1/64^{th}$ of the size of physical memory in MB, just like `ncsize` and `ufs_ninode`. For example, a Solaris NFS client with 64 MB of RAM will only be able to keep rnode information about 1144 files it has opened, unless the value of `nrnode` is increased. There is no need to increase `nrnode` on a server. We have increased the value of `ncsize` from 2232 to 12288.

The size of the UFS metadata buffer cache `bufhwm` (in kB), which defaults to 2% of physical memory, had to be increased as well, since the buffer has to be big enough to hold all active and inactive inodes (approx. 300 bytes per inode) and DNLC cache entries (approx. 30 bytes per entry). If the server is serving only very small files, there can as many as file cache size/page size $\approx$ 80 MB / 4 kB = 20000 files in the cache and active inodes in the buffer cache (of course, such a situation looks too artificial).

Another fraction of a percent improvement in performance can be achieved by reducing the rate at which `fsflush` scans the memory looking for pages belonging to modified files that have to be written to the disk. Since all NFS writes are synchronous, the `autoup` time can be safely increased from 30 seconds to 1 minute.

| Parameter name | Default size[21] | New size | Line to add to `/etc/system` |
|---|---|---|---|
| `lotsfree` | 512 | 2048 | `set lotsfree=2048` |
| `slowscan` | 100 | 100 | `set slowscan=100` |
| `fastscan` | 8192 | 8192 | `set fastscan=8192` |
| `ufs_ninode` | 2232 | 12288 | `set usf_ninode=12288` |
| `ncsize` | 2232 | 12288 | `set ncsize=12288` |
| `bufhwm` | 2621 | 12288 | `set bufhwm=12288` |
| `nrnode` | 2232 | 2232 | `set nrnode=2232` |
| `autoup` | 30 | 60 | `set autoup=60` |
| `tune_t_fsflush` | 5 | 10 | `set tune:tune_t_fsflush=10` |
| `maxpgio` | 40 | 60 | `set maxpgio=60` |

---

[21] See Chapter 69, "Overview of System Performance", in the *Solaris System Administration Guide, Volume 2*.

## 6. Performance testing and results

We tested the server's performance on several distinctively different types of files:

- very large files (2 files 137.6 MB each). These files are larger than the amount of memory available to the file cache on our Solaris x86 server, from now on denoted by **S**, so that by the time the second file has been read there is no trace of the first file left in the file cache. This is essentially a test of sequential I/O. Since some NFS clients used in testing have several hundred megabytes of RAM, additional precautions must be taken to avoid caching on the client side.

- large files (24.6 MB, a release 4.2.1 `BetaApp` executable). Files of this size are expected to be fully cached both on the client and the server sides. As has been mentioned in the previous chapter, Solaris x86 has problems with it, but it does not really matter.

- very many very small files (typical BaBar source code: approximately 28,000 `.c`, `.cc`, `.h`, `.hh`, `.F`, etc. files, average size 4.7 kB, total size 137.6 MB when `tar`red, a part of the `$BFDIST/packages` tree). Even our super-sized inode cache cannot hold information about this many files. Most NFS clients will run out of cache space as well.

- Not that many very small files (approximately 7530 files, average size 4.7 kB, total size 35 MB when `tar`red). These files are expected to be fully cached both on the client and the server sides, unless the client runs out of space in its rnode cache.

Precautions have been taken to avoid undesired file caching both on the server and the client. Most tests were conducted at night or on weekends and repeated at least twice.

The etalon (**A**), which our new server (**S**) is compared with throughout this chapter, is a $100k+ Digital's AlphaServer 2100 5/250[22] running Digital UNIX 4.0, complete with four Alpha 21164 250 MHz CPUs, 640 MB of RAM, and connected to 100 Mb/s Fast Ethernet network (100Base-TX, half-duplex mode, on one subnet with **S**). All disks on **A** are 3.5" wide UltraSCSI (40 MB/s interface bandwidth)[23] hooked up through a RAID controller. Some logical disks on **A** are RAID stripes made of partitions located on 3 different physical drives, others consist of a single partition of a single physical disk. Unless noted otherwise, the latter were used in all tests. **A** is the software development platform and the primary file server for the LBNL BaBar group.

Other machines mentioned in this chapter are (disk type information unavailable):

- **B**: one 333 MHz AlphaStation, 320 MB RAM, Digital UNIX 4.0; 10Base-T;
- **C**: six 50 MHz SPARC processors, 192 MB RAM, Solaris 2.5.1, 100Base-TX;

We will provide CPU load information for most performance tests. The purpose is to give the reader an idea how CPU-intensive disk or network I/O is, depending on the type of files, the power and number of CPUs. In most cases, we used `iostat`. On Solaris, it reports the percentage of time the processor spends in the system mode (`sy`, corresponds to `%sys` in `sar -u`), in the user mode (`us`, `%user`), idle and waiting for *disk* I/O completion (`wt`, `%wio`), and idle and not waiting for I/O (`id`, `%idle`), as well as disk and terminal I/O. On Digital UNIX, `iostat` does not report the disk I/O wait time, including it in the CPU idle time, so we used a utility called `uaio`[24], which is is an improved version of `iostat` for Digital UNIX. The `top` utility was used to monitor activity of other processes in order to make sure that they do not significantly interfere with our benchmarks (the CPU utilization statistics reported by `top` has been found to be inconsistent with statistics reported by other utilities, so we do not use it as a system performance monitor). `/usr/bin/time` was used for timing; all times are in seconds. We keep more digits than are statistically significant.

---

[22] See `http://www.digital.com/alphaserver/archive/2100/alphaserver2100.html`

[23] See `http://www.storage.digital.com/swrks/guide/pg_ddrv.htm`.

### Test #1: Local 137.6 MB file read (/usr/bin/time tar −cvf /dev/null verybigfile)

The file is read two times in a row to see the filesystem cache performance. Cache-influenced 'bogus' I/O rates are given in *italic*. Reminder: **A** has 4 CPUs, so a load of 25% means that one of the CPUs is 100% busy. We use `tar` because `cp` is too smart: `cp file /dev/null` completes immediately; `wc file` or `more file > /dev/null` spend more time in the `user` mode than `tar`.

| MB/s | **S**: exported directory | | | | | | | | **A**: scratch directory | | | | | | | |
|---|---|---|---|---|---|---|---|---|---|---|---|---|---|---|---|---|
| | I/O | real | user | sys | %user | %sys | %wio | %idle | I/O | real | user | sys | %user | %sys | %wio | %idle |
| first read | **3.63** | 37.9 | 0.2 | 9.0 | 1 | 96 | 0 | 3 | **3.54** | 38.8 | 0.0 | 3.0 | 1 | 4 | 0 | 95 |
| second read | **3.60** | 38.2 | 0.3 | 8.8 | 1 | 95 | 0 | 4 | *51* | 2.7 | 0.0 | 2.7 | 1 | 26 | 0 | 73 |

### Test #2: Local 24.6 MB file read (/usr/bin/time tar −cvf /dev/null bigfile)

You can see the weird behavior of the file system cache on **S** mentioned in the previous chapter. A non-zero value of `%wio` indicates non-sequential I/O, i.e. fragments of this file in the filesystem cache were recycled by the system before the file was read again.

| MB/s | **S**: exported directory | | | | | | | | **A**: scratch directory | | | | | | | |
|---|---|---|---|---|---|---|---|---|---|---|---|---|---|---|---|---|
| | I/O | real | user | sys | %user | %sys | %wio | %idle | I/O | real | user | sys | %user | %sys | %wio | %idle |
| first read | **3.72** | 6.6 | 0.0 | 1.6 | 1 | 99 | 0 | 0 | **3.51** | 7.0 | 0.0 | 0.6 | 0 | 4 | 1 | 95 |
| second read | *7.24* | 2.9 | 0.0 | 1.2 | 1 | 80 | 20 | 1 | *49* | 0.5 | 0.0 | 0.5 | 1 | 27 | 0 | 72 |
| third read | *17.6* | 1.2 | 0.0 | 1.2 | 5 | 95 | 0 | 0 | *49* | 0.5 | 0.0 | 0.5 | 0 | 26 | 1 | 73 |

### Test #3: Local 28,000 file read (/usr/bin/time tar −cvf /dev/null ./zillion/ >/dev/null)

In this test, I/O rate is measured in files per second. As we can see, **A** is not capable of caching this many tiny files: it runs out of space in the inode cache. One of the CPUs on **A** spends about 70% of time in the I/O wait mode.

| files/s | **S**: exported directory | | | | | | | | **A**: scratch directory | | | | | | | |
|---|---|---|---|---|---|---|---|---|---|---|---|---|---|---|---|---|
| | I/O | real | user | sys | %user | %sys | %wio | %idle | I/O | real | user | sys | %user | %sys | %wio | %idle |
| first read | **154** | 180 | 11.9 | 37.1 | 7 | 42 | 46 | 5 | **156** | 179 | 5.6 | 25.5 | 2 | 6 | 19 | 74 |
| second read | **156** | 179 | 12.5 | 36.7 | 6 | 41 | 47 | 6 | **159** | 175 | 5.6 | 26.1 | 2 | 5 | 18 | 75 |

### Test #4: Local 28,000 file write (/usr/bin/time tar −xvf ../verybigfile.tar >/dev/null)

---

[24] The `uaio` utility was written by Kurt Carlson at the University of Alaska. The latest version can be downloaded from `ftp://raven.alaska.edu/pub/sois/`.

This test includes reading `verybigfile.tar` from the disk and `untar`ring it into an empty directory. The write cache affects this test a little bit (the data get written to the disk only after `fsflush` discovers that pages in the memory associated with open files have been modified). `%sys` is larger in this test than in the previous one because the CPU has to think how to allocate the disk space for the files that are being written. For some reason, **S**'s score is much better than **A**'s. This is a big surprise.

| files/s | **S**: exported directory | | | | | | | | **A**: scratch directory | | | | | | | |
|---|---|---|---|---|---|---|---|---|---|---|---|---|---|---|---|---|
| | I/O | real | user | sys | %user | %sys | %wio | %idle | I/O | real | user | sys | %user | %sys | %wio | %idle |
| write | **64** | 436 | 13.7 | 120 | 3 | 51 | 45 | 1 | **27** | 1043 | 4.7 | 37.2 | 1 | 3 | 25 | 71 |

## Test #5: Local 28,000 file remove (/usr/bin/time rm −r *)

We are now removing 28,000 tiny files that we have just written to the disk. There is some cheating involved in this test, since **S** still has access information about most of these files in its DNLC and inode caches. The purpose of this test is to demonstrate the power of these caches.

| files/s | **S**: exported directory | | | | | | | | **A**: scratch directory | | | | | | | |
|---|---|---|---|---|---|---|---|---|---|---|---|---|---|---|---|---|
| | I/O | real | user | sys | %user | %sys | %wio | %idle | I/O | real | user | sys | %user | %sys | %wio | %idle |
| write | **133** | 210 | 2.8 | 55.7 | 1 | 43 | 56 | 0 | **41** | 683 | 1.9 | 23.7 | 1 | 2 | 24 | 73 |

## Test #6: Local 7530 file read (/usr/bin/time tar −cvf /dev/null ./xillion/ >/dev/null)

Now **S** is able to cache both the files themselves and their inodes. Before the first read in the first table, the files' access information *was not* in the inode and DNLC caches, and the inode cache was full. Before the first read in the second table, it *was* cached (we purged all data belonging to those files from the file cache by reading two 137.6 MB files; this procedure left file access information in the inode and DNLC caches intact). 7530 files is still too many for the **A**'s file access cache, but it does do some work.

| files/s | **S**: exported directory | | | | | | | | **A**: scratch directory | | | | | | | |
|---|---|---|---|---|---|---|---|---|---|---|---|---|---|---|---|---|
| | I/O | real | user | sys | %user | %sys | %wio | %idle | I/O | real | user | sys | %user | %sys | %wio | %idle |
| first read | **153** | 49.1 | 5.0 | 10.3 | 3 | 38 | 57 | 0 | **149** | 50.4 | 1.5 | 6.6 | 1 | 4 | 20 | 76 |
| second read | *593* | 12.7 | 3.9 | 6.7 | 16 | 84 | 0 | 0 | *191* | 39.4 | 1.6 | 6.2 | 1 | 5 | 19 | 75 |
| third read | *753* | 10.0 | 3.5 | 5.8 | 38 | 38 | 0 | 0 | *224* | 33.5 | 1.5 | 5.9 | 1 | 5 | 19 | 75 |

| files/s | **S**: exported directory | | | | | | | | **A**: scratch directory | | | | | | | |
|---|---|---|---|---|---|---|---|---|---|---|---|---|---|---|---|---|
| | I/O | real | user | sys | %user | %sys | %wio | %idle | I/O | real | user | sys | %user | %sys | %wio | %idle |
| first read | **187** | 40.3 | 3.3 | 8.4 | 11 | 36 | 51 | 1 | | | | | | | | |
| second read | *655* | 11.5 | 3.6 | 6.0 | 16 | 84 | 0 | 0 | | | Almost same as above | | | | | |
| third read | *746* | 10.1 | 3.9 | 5.6 | 38 | 38 | 0 | 0 | | | | | | | | |

## Test #7: NFS 137.6 MB file read
## (/usr/bin/time tar −cvf /dev/null /home/server/verybigfile)

We will skip tests of the NFS clients' ability to cache remote files. We will only test NFS servers' performance. In all cases, the file was not in the server's file cache before the test. Multiprocessor machines are marked with *. Transfers between two machines connected to 100 Mb/s Ethernet are denoted by "=>"; if any of the two is on 10 Mb/s Ethernet, then by "->". As we can see, server throughput over 100 Mb/s Ethernet is higher than over 10 Mb/s Ethernet, but the effect is not as large as one might expect. It is also seen that the **S**'s performance in this test is CPU-bound and could be better if it had a faster or a dual processor and/or SCSI disks instead of EIDE.

| NFS read, MB/s | I/O | client | | | | | | | server | | | |
|---|---|---|---|---|---|---|---|---|---|---|---|---|
| | | real | user | sys | %user | %sys | %wio | %idle | %user | %sys | %wio | %idle |
| **S** => **A**\* | **1.90** | 72.4 | 0.1 | 4.9 | 1 | 8 | 0 | 91 | 0 | 100 | 0 | 0 |
| **S** => **C**\* | **1.88** | 73.0 | 0.4 | 15.7 | 1 | 11 | 0 | 89 | 0 | 98 | 0 | 2 |
| **A**\*† => **C**\* | **1.95** | 70.6 | 0.5 | 14.9 | 1 | 14 | 0 | 85 | 1 | 6 | 8 | 85 |
| **S** -> **B** | **0.94** | 146.1 | 0.1 | 1.5 | 0 | 5 | 0 | 95 | 0 | 49 | 0 | 51 |
| **A**\* -> **B** | **1.04** | 132.0 | 0.1 | 1.7 | 0 | 6 | 0 | 94 | 1 | 3 | 0 | 96 |

`† stripe`

### Test #8: NFS 24.6 MB file write (/usr/bin/time tar −cvf /home/server/bigfile bigfile)

The file is loaded into the NFS client's file cache before the test, so it is a pure write test. All NFS writes are synchronous, i.e. the system must make space allocation decisions as the data is being received. and that is one of the reasons why NFS writes are slower than NFS reads. Just like in test #7, using 100 Mb/s Ethernet has no dramatic effect on throughput.

| NFS write, MB/s | I/O | client | | | | | | | server | | | |
|---|---|---|---|---|---|---|---|---|---|---|---|---|
| | | real | user | sys | %user | %sys | %wio | %idle | %user | %sys | %wio | %idle |
| **S** <= **A**\* | **1.68** | 14.6 | 0.0 | 2.6 | 1 | 11 | 0 | 88 | 0 | 93 | 0 | 7 |
| **S** <= **C**\* | **1.36** | 18.1 | 0.0 | 8.3 | 1 | 11 | 66 | 21 | 0 | 72 | 0 | 27 |
| **A**\*† <= **C**\* | **1.57** | 15.7 | 0.1 | 5.3 | 1 | 16 | 0 | 83 | 1 | 11 | 35 | 53 |
| **S** <- **B** | **0.95** | 25.8 | 0.0 | 0.8 | 0 | 6 | 9 | 94 | 0 | 55 | 0 | 45 |
| **A**\* <- **B** | **0.56** | 44.3 | 0.0 | 1.6 | 0 | 7 | 0 | 93 | 0 | 4 | 0 | 96 |

`† stripe`

### Test #9: NFS 7530 small files file read
### (/usr/bin/time tar −cvf /dev/null /home/server/xillion)

Neither the files nor their access information were present in server's memory before the test. If the files' inodes and directory information were already in RAM, results would be better. Comparing these results with local reads (Test #6), we see that the server spends a larger fraction of non-idle time in the system mode, waiting for completion of network I/O.

| NFS read, files/s | I/O | Client | | | | | | | server | | | |
|---|---|---|---|---|---|---|---|---|---|---|---|---|
| | | real | user | sys | %user | %sys | %wio | %idle | %user | %sys | %wio | %idle |
| **S** => **A**\* | **99** | 76.2 | 1.6 | 11.2 | 1 | 6 | 0 | 93 | 0 | 40 | 36 | 24 |
| **S** => **C**\* | **71** | 106.3 | 1.4 | 6.5 | 2 | 20 | 0 | 78 | 0 | 32 | 17 | 52 |
| **A**\* => **C**\* | **66** | 114.2 | 1.2 | 6.4 | 2 | 17 | 0 | 81 | 0 | 5 | 13 | 82 |
| **S** -> **B** | **71** | 105.4 | 1.3 | 9.3 | 2 | 16 | 0 | 82 | 0 | 31 | 24 | 45 |
| **A**\* -> **B** | **65** | 115.3 | 1.2 | 9.2 | 2 | 15 | 0 | 83 | 1 | 4 | 14 | 82 |

### Test #10: NFS 7530 small files file write
### (/usr/bin/time tar −xvf xillion.tar )

Opening a file for writing over NFS takes more time opening a file for reading. This overhead together with effects discussed earlier (see tests #4 and #8) make NFS writes especially slow for small files. The poor performance of **A**'s disks still remains a mystery.

| NFS write, files/s | I/O | client | | | | | | | server | | | |
|---|---|---|---|---|---|---|---|---|---|---|---|---|
| | | real | user | sys | %user | %sys | %wio | %idle | %user | %sys | %wio | %idle |
| **S <= A\*** | **40.4** | 186.4 | 1.3 | 16.1 | 1 | 5 | 0 | 94 | 0 | 71 | 14 | 15 |
| **S <= C\*‡** | **24.9** | 303.0 | 8.4 | 59.7 | 1 | 7 | 3 | 90 | 0 | 38 | 47 | 15 |
| **A\* <= C\*‡** | **5.5** | 1362 | 11.9 | 594 | 1 | 12 | 22 | 66 | 0 | 1 | 7 | 92 |
| **A\*† <= C\*‡** | **8.9** | 849.5 | 9.7 | 151.3 | 1 | 11 | 29 | 58 | 0 | 1 | 10 | 89 |
| **S <- B** | **35.5** | 212.1 | 0.8 | 8.5 | 0 | 6 | 0 | 94 | 0 | 60 | 16 | 24 |
| **A\* <- B** | **13.2** | 570.6 | 0.8 | 8.6 | 0 | 3 | 0 | 97 | 1 | 2 | 19 | 78 |

† stripe          ‡ upgraded to Solaris 2.6

## 7. Conclusion

We have approached a standard problem in a truly scientific way — by looking for a non-standard solution. The pilot project is a success: from a $2,100 ($1,500 in February 1998) worth of hardware, we have built an NFS file server with 19.2 GB of disks, expandable to 112 GB for another $6,200. For a fraction of the price, even in its current configuration with relatively slow CPU and hard drives, it has proven to be a worthy contender in performance to its proprietary counterparts. If necessary, its performance can be improved by upgrading to a faster or a dual CPU. We also cannot say anything negative about its reliability. The server has been in use for four months and so far has not exhibited any software or hardware problems or anomalous behavior. As of the moment this note is being submitted to the BaBar Note repository, it has been up for 58 days.

Solaris x86 is an NFS file server platform well suited for the needs of a medium-size experimental particle physics workgroup. We recommend it for use.

## Acknowledgements

I would like to thank Mike Bordua of the LBNL's Computing Sciences Division for valuable discussions and suggestions regarding Solaris setup. I would also like to thank Tom Pavel (SLAC) for comments on various aspects determining server's performance.